\documentclass[10pt,a4paper,onecolumn]{article}

\usepackage{microtype}
\usepackage{science}
\usepackage{amsmath}
\usepackage{amssymb}
\usepackage{wasysym}
\usepackage[italian,english]{babel}
\usepackage{amsthm}
\usepackage{amscd}
\pdfoutput=1 

\usepackage[colorlinks,hyperindex,plainpages=false]{hyperref}
\usepackage{hypcap}
\usepackage{thumbpdf}
\usepackage{url}

\title{\center\Large\bf Nonstandard Finite Difference Scheme for Mutualistic Interaction Description}

\author{Gianluca Gabbriellini}

\begin{document}
\maketitle

\begin{abstract}
One of the more interesting themes of the mathematical ecology is the description of the mutualistic interaction between two interacting species. Based on continuous--time model developed by Holland and DeAngelis 2009 for consumer--resource mutualism description, this work deals with the application of the Mickens Nonstandard Finite Difference method to transform the continuous--time scheme into a discrete--time one. It has been proved that the Mickens scheme is dynamically consistent with the original one regardless of the step-sizes used in numerical simulations, in opposition of the Forward--Euler method that shows numerical instabilities when the step size overcomes a critical value. 

\end{abstract}

\section{Introduction}

The mutualism is acknowledged as a fundamental process in ecological systems: it is an interaction between individuals of different species, from which the interacting populations have beneficial effects. Like predation and competition, is now recognized as a consumer--resource (C--R) interaction \cite{Holland_deangelis2010}. The C--R interaction relates the process of nutrient transfer between a consumer and a resource: is a mechanism that describes the way by which individuals interact with each other \cite{Holland_deangelis2009}. All the types of mutualism belong to one of the three general forms of C--R interaction: uni--directional, bi--directional and indirect C--R mutualisms \cite{Holland_2005}. 

From a mathematical point of view only few authors have been attempts to modelize the mutualism, probably because it is difficult to retrieve the experimental data. One of the first attempts is based on Lotka--Volterra competition equations \cite{may}, but it conducts to the boundless growth of the interacting population. Subsequently, other models are proposed where a stable equilibrium for populations are reached \cite{soberon}, \cite{dean83}. Holland and DeAngelis \cite{Holland_deangelis2009} developed a general continuous--time model to describe the population dynamic for two mutualistic species; whose model is based on a modified version of the Rosenzweig--MacArthur \cite{rosenz} model of predation.

Although a continuous--time model based on differential equations constitutes the first step to modelize a population evolution problem, for the more complexes issues investigated in mathematical biology, the numerical method is the only practicable way. Also, discrete--time models become essential when one wants to describe experimental data that have been collected with a certain interval of time. One of the critical aspects of the discretization methods is just the dependence from this interval, the time step. Since the time step should be selected only in relation to the characteristics of the problem under examination, it is necessary to choose a reliable discretization method that allows to make this transformation without any restriction on time step, as well as not to introduce artifacts, change the linear stability properties etc.. Since, in general, these requirements are not satisfied from all the discretization methods (see i.e., \cite{mickens01}), in this paper is presented the application of the Nonstandard Finite Difference scheme \cite{mickens89} to the system of differential equations discussed by Holland and DeAngelis in the articles \cite{Holland_2005} and \cite{Holland_deangelis2010}. In this work are only considered the equations describing the population dynamic for the bi--directional C--R mutualism since the uni--directional problem is easily solved once the bi--directional one is solved.

\section{Time Continuous Approach}

A  mathematical model for population growth can be generally described by a system of first order of \textit{autonomous ordinary differential equations}:
\begin{align}\label{eq:differential_equation}
\left\{
\begin{array}{l}
\frac{d\xi}{dt}=f(\xi,k) \\ \\
\xi(t_0)=\xi_0,  \\
\end{array}
\right.
\end{align}
with $f: \mathbb{R}^n\longrightarrow \mathbb{R}^n$ is supposed to be sufficiently smoothed; $\xi=\xi(t):[0,+\infty)\longrightarrow \mathbb{R}^n$ are the coordinates, with initial condition $\xi_0\in\mathbb{R}^n$; finally $k=(k_1,k_2,...)$ represents the system parameters. If $f$ is globally Lipschitz then there exists a unique solution $\xi$ for all $t\ge t_0$, hence the eq. \eqref{eq:differential_equation} defines a dynamical system on $\mathbb{R}^n$. For a system defined with eq. \eqref{eq:differential_equation} we define a steady state solution (or also fixed point of $f$) a point $\tilde\xi\in\mathbb{R}^n$ that satisfies the relation:
\begin{align}\label{eq:fixed_point_cont}
f(\tilde\xi)=0, 
\end{align}
In the continuation of the text will be used the set $\Gamma_c=\{\tilde\xi|f(\tilde\xi)=0,\,\tilde\xi\in\mathbb{R}^n\}$ to refer to the set of steady states solutions.

If $f\in C^1$ it is possible to calculate the $n\times n$ Jacobian matrix as here indicated with $J_f$. In term of the Jacobian matrix, as given $\sigma(J_f)$ the set of its eigenvalues, a theorem ensures that a steady--state $\tilde\xi\in\mathbb{R}^n$ for which $J$ has no eigenvalues with zero real parts (hyperbolic steady--state) is \cite{lyapunov1992}:
\begin{itemize}
\item \textit{asymptotically stable} if and only if $\Re\lambda<0$, for all $\lambda\in\sigma(J_f)$,
\item \textit{unstable} if and only if $\Re\lambda>0$, for all $\lambda\in\sigma(J_f)$. 
\end{itemize}
The same is not true for a non--hyperbolic steady--state.

\section{Finite Difference Approaches}
To transform a continuous--time model into a discrete one, the continuous variable $t\in [0,\infty)$ must be replaced by the discrete variable $n\in\,\mathbb{N}$ and the variable $\xi$ must take discrete values $\xi_n$. The result is a difference equation. Let $f:\mathbb{R}^n\longrightarrow\mathbb{R}^n$, consider a sequence $\{\xi_n\}_{n=0}^{\infty}$: it can be defined by a mapping $~\Lambda:\mathbb{R}^n\times\mathbb{R}^n\longrightarrow \mathbb{R}^n$ of the form $H(\xi_{n+1},\xi_{n})$. In some case, such as the one discussed on this paper, is possible that $\xi_{n+1}$ is given explicitly in terms of $\xi_{n}$:
\begin{align}\label{eq:diff_eq0}
\xi_{n+1}=F(\xi_n),
\end{align}
where $F:A\subseteq\mathbb{R}^n\longrightarrow\mathbb{R}^n$.
 
For a dynamical discrete system defined by eq. \eqref{eq:diff_eq0} a steady--state $\tilde\xi_n\in\mathbb{R}^n$ respects the following conditions:
\begin{align}\label{eq:fixed_point_disc}
F(\tilde\xi_n)=\tilde\xi_n.
\end{align}
Likewise to the continuous case, also in discrete case it is useful to define with $\Gamma_d=\{\tilde\xi_n|F(\tilde\xi_n)=0,\,\tilde\xi_n\in\mathbb{R}^n\}$ the set of steady--states. 

To check the stability of the system described by eq. \eqref{eq:fixed_point_disc} the procedure is similar to the continuous case. If $F\in C^1$ is possible to calculate the $n\times n$ Jacobian matrix, indicated with $J_{F}$. Given $\sigma(J_F)$ the set of the Jacobian eigenvalues, a theorem (see i.e., \cite{mathmodecol}) ensures that a steady--state $\tilde\xi_n\in\mathbb{R}^n$ is:
\begin{itemize}
\item locally \textit{asymptotically stable} if and only if $\Re\lambda<1$ $\forall\,\lambda\in\sigma(J_F)$: this point is an \textit{attractor}.
\item \textit{unstable} if and only if $\Re\lambda>1$ $\forall\lambda\in\sigma(J_F)$: this point is a \textit{repeller}.
\item no conclusions on stability if $\Re\lambda>1$ for some $\lambda\in\sigma(J_F)$. 
\end{itemize}
The finite difference method is called \textit{elementary stable} if, for any value of the time step $\Delta t$, the linear stability of each $\tilde\xi_n\in\Gamma_d$ is the same as the stability of each $\tilde\xi\in\Gamma_c$.

From a theoretical point of view this criterion is good in every case to control the stability of the steady--states. Nevertheless, in the case of multi--dimensional systems, expecially in two and three dimensions, it is more convenient to use an alternative method, the so--called Jury criterion \cite{Jury}. Let $\tilde J_F:= J_F(\tilde \xi_n)$, for two dimensional systems the characteristic polynomial of the Jacobian can be written as:
\begin{align}
\lambda^2-\lambda \operatorname{tr} (\tilde J_F)+\det(\tilde J_F).
\end{align}
In order to have $\Re\lambda<1$ for all $\lambda\in\sigma(J_F)$, the Jury condition states that:
\begin{align}\label{eq:Jury}
|\operatorname{tr}(\tilde J_F)|<1+\det(\tilde J_F)<2.
\end{align}
Therefore, this criterion establishes that exists a necessary and sufficient condition to guarantee the asymptotical stability of the steady--state solutions. 

\paragraph{The Forward--Euler Method.} This is one of the oldest way to derive a finite difference equation from a differential equation. This numerical procedure requires that the system \eqref{eq:differential_equation} is transformed by introducing these substitutions:
\begin{align}\label{eq:femrule1}
\xi(t)\longrightarrow \xi_n,
\end{align}
\begin{align}\label{eq:passaggiocont-disc}
\frac{d\xi(t)}{dt}\longrightarrow \frac{\xi_{n+1}-\xi_n}{\Delta t},
\end{align}
where $\Delta t$ is the step size and $\xi_{n+1}\approx \xi(t+n\Delta t)$. As showed by Mickens (see i.e., the reference \cite{mickens02}), in many cases the transformation of eq. \eqref{eq:passaggiocont-disc} leads to numerical instabilities which occur as solutions of the finite difference equations, steady--states, bifurcations etc. that do not appear in the originary differential equation. 

\subsection{Nonstandard Finite Difference method}  
To avoid these potential problems, Mickens \cite{mickens89} suggests what is known as the Nonstandard Finite Difference (NSFD) method. An important concept on which the NSFD schemes is based is the Dynamic Consistency \cite{mickens05}. This criterion states that, let the first--order autonomous ODE 
\begin{align}\label{eq:diff_eq}
\frac{d\xi}{dt}=f(\xi,k),
\end{align}
and given $U$ its set of properties, the difference equation
\begin{align}
\xi_{n+1}=F(\xi_n,k,\Delta t),
\end{align}
is \textit{dynamically consistent} with \eqref{eq:diff_eq} if it respects the same set of properties. As indicated in \cite{alkahby}, the dynamic consistency is verified when a difference equation possesses the same stability, bifurcations and chaos of the original differential equation.

A finite difference method is called a NSFD scheme is almost one of the following conditions is satisfied:
\begin{enumerate}
\item[I.] The order of the discrete derivative should be equal to the order of the corresponding derivatives of the differential equation. 
\item[II.] Denominator functions for the discrete rapresentation must be non--trivial. The following replacement is then required: 
\begin{align*}\Delta t\longrightarrow\phi(\Delta t)+O(\Delta t^2),\end{align*}
where $\phi(\Delta t)$ is such that $0<\phi(\Delta t)<1$, $\forall\Delta t>0$.
\item[III.] Nonlinear terms must be replaced by non--local discrete representations, i.e.,
\begin{align*}\xi^2\longrightarrow\xi_n\xi_{n+1}.\end{align*}
\item[IV.] Special conditions that hold for the solutions of the differential equations should also hold for the solutions of the finite difference scheme.
\item[V.] The scheme should not introduces spurious solutions.
\end{enumerate}
The NSFD scheme incorporates the principle of Dynamical Consistency.

A very important characteristic of dynamical systems, especially in those of biological interest, is that all solutions must remain non--negative in order to maintain the problem well--posed, from biological and mathematical points of view. A method that respects the Mickens rules (I to V) and preserves the solution positivity is called Positive and Elementary Stable Nonstandard (PESN) method. The appropriate non--local approximation that provides a such scheme is constructed by using the indications suggested by Patankar \cite{patankar}. Applying these expedients, the discrete scheme will comply with the physical properties of the differential equations, without any restriction on the step size $\Delta t$.

\section{An Elementary Application: the Logistic Population Model}
The logistic equation, originally due to Pierre Fran\c{c}ois Verhulst \cite{verhulst}, is the law that regulates with good approximation the growth rate of a certain population number as function of time. The model is based on the assumption that the population evolves in an environment with limited resources with no immigration or emigration phenomena. 
Let $x(t)$ the population at instant $t$, the law that regulates it can be expressed from the following first--order autonomous ODE:
\begin{align}\label{eq:logistic}
\frac{dx(t)}{dt}=rx(t)-\frac{r x^2(t)}{E},
\end{align}
where $E$ is the \textit{carrying capacity} of the system and $r$ is the \textit{intrinsic growth rate} ($r=b-d$, where $b$ is the instantaneous birth rate and $d$ the instantaneous death rate); also we assume $r,\,k>0$. The eq. \eqref{eq:logistic} is the same as \eqref{eq:differential_equation} assuming $\xi=x(t)$ and $f$ coinciding with the right hand term. In this equation the term $rx(t)$ is the intrinsic population growth rate, while $-rx^2(t)/E$ is the self--limitating term.
These two factors of $f$ have opposite sign then $f$ can be greater, lesser or equal to zero: this determines respectively the increase, the decrease or the constancy of the population.
 
Applying the condition dicted by eq. \eqref{eq:fixed_point_cont}, is simple to calculate the two steady--states: a trivial solution $\tilde x_1=0$ and a non--trivial $\tilde x_2=E$. Substituting these solutions in the Jacobian $J(x)=1-2x/E$ the following inequalities $J(\tilde x_1)>0$ and $J(\tilde x_2)<0$ hold. We can deduce that only the solution $\tilde x_2=E$ is asymptotically stable. 
\\This can be deduced also by an alternative way. The differential equation \eqref{eq:logistic}, solved by variables separation technique, admits the following solution:
\begin{align}\label{eq:logist_sol}
x(t)=\frac{E}{1+\big(\frac{E}{x_0}-1\big)e^{-rt}}.
\end{align}
Calculating the limit of eq. \eqref{eq:logist_sol} for $t\longrightarrow +\infty$ the population has just $E$ as its asymptotic value.

\subsection{Discrete model of logistic equation.} Following the NSFD rules, the logistic model of eq. \eqref{eq:logistic} can be transformed in the discrete scheme:
\begin{align}\label{eq:logistic_disc}
\frac{\Delta x(t)}{\phi(\Delta t)}=rx(t)-\frac{r x^2(t)}{E},
\end{align}
where $\Delta x(t)=x_{n+1}-x_n$ and $\phi(\Delta t)$ is a function of the time step in respect of the rule II; in accord with rule III, $x^2\longrightarrow x_nx_{n+1}$. Therefore $x_{n+1}$ can be so explicited:
\begin{align}\label{eq:logistic_disc1}
x_{n+1}=\frac{1+r\phi(\Delta t)}{1+\frac{r\phi(\Delta t)}{E}x_n}x_n,
\end{align}
where $x_0=x(t=0)$ is the initial population. It is simple to show that the condition of positiveness $x_{n+1}\ge 0$ is automatically satisfied if $\phi$ satisfy the rule II of the NSFD scheme. The steady--states of eq. \eqref{eq:logistic_disc1} are $\tilde x_1=0$ and $\tilde x_1=E$: for the trivial steady--state, $J(\tilde x_1)=1+r\phi>1$ for $ r>0$ and $ 0<\phi<1$, then is unstable; for the non trivial one, $J(\tilde x_2)=1/(1+r\phi)<1$, then is asimptotically stable. We conclude that the NSFD scheme well reproduces the behavior of the differential equation \eqref{eq:logistic_disc}.

In opposition, the Euler and Runge--Kutta discretization methods are dynamically inconsistent with the original differential equation \cite{mickens94}, showing numerical instabilities. Also in \cite{anguelov_lubuma_shillor} is remarked that the application of these two standard schemes produces a not correct monotonicity and oscillations for particular values of the mesh $\Delta t$.

\section{Mutualistic Interaction Between two Species} 

It is now considered a more complex case respect to the logistic equation: this is the formulation introduced by Holland and DeAngelis 2009 \cite{Holland_2005} for the mutualism description, based on Rosenzweig--MacArthur model of predator--prey interactions. After a brief review of the problem in continuous--time domain, two discrete schemes will be proposed. 

\subsection{Continuous--time domain}\label{subsec:ctd}

Subsequently it will be denoted with $x(t)$ and $y(t)$ the numbers (or densities) of the mutualistic species. In a mutualistic model, the growth rate of each species must involves the same terms of the logistic equation \eqref{eq:logistic} that represent effect separate from mutualism: a linear term that represent the intrinsic population growth rate and a quadratic one to modify the growth rate with density dependent self--limitation. Also, for each species, in case of bi--directional mutualism, other two terms are required:
\begin{itemize}
\item a positive term that quantify the advantages that the presence of a species induces on the growth of the other ($\alpha_{ij}$ in eq. \eqref{eq:mutualism_timecontinuos}),
\item a negative term that describes the disadvantages that the presence of a species induces on the growth of the other ($\beta_{i}$ in eq. \eqref{eq:mutualism_timecontinuos}). 
\end{itemize} 
With these two conditions, a community of two mutualistic species can be described by the following system \cite{Holland_2005}:
\begin{equation}\label{eq:mutualism_timecontinuos}
\left\{
\begin{array}{l}
\frac{dx}{dt}=r_{1}x+\alpha_{12}\frac{xy}{h_2+y}-\beta_1\frac{xy}{e_1+x}-d_1x^2\\ \\

\frac{dy}{dt}=r_{2}y+\alpha_{21}\frac{xy}{h_1+x}-\beta_2\frac{xy}{e_2+y}-d_2y^2\\ \\

x(0)\ge 0,\,y(0)\ge 0, \\
\end{array}
\right.
\end{equation}
where all parameters are constants, real and non--negative. To create correspondence with the notation of eq. \eqref{eq:differential_equation}, we assume $k_1=(r_1,\,\alpha_{12},\,h_2,\,\beta_1,\,e_1,\,d_1)$ and $k_2=(r_2,\,\alpha_{21},\,h_1,\,\beta_2,\,e_2,\,d_2)$.

\subsection{Discrete--time domain}\label{sub:dtd}

\paragraph{Forward--Euler method.} It is required to apply the rules \eqref{eq:femrule1} and \eqref{eq:passaggiocont-disc} to the differential equation system \eqref{eq:mutualism_timecontinuos}. The result is:
\begin{equation}\label{eq:mutualism_timedisc_fem}
\left\{
\begin{array}{l}
\frac{x_{n+1}-x_n}{\Delta t}=r_{1}x_n+\alpha_{12}\frac{y_nx_n}{h_2+y_n}-\beta_1\frac{y_nx_{n}}{e_1+x_n}-d_1x_n^2\\ \\

\frac{y_{n+1}-y_n}{\Delta t}=r_2y_{n}+\alpha_{21}\frac{y_nx_n}{h_1+x_n}-\beta_2\frac{y_{n}x_n}{e_2+y_n}-d_2y_n^2\\ \\

x_0>0, y_0>0, \\
\end{array}
\right.
\end{equation}
that can be so explicited:
\begin{equation}\label{eq:mutualism_timedisc_sol_eul}
\left\{
\begin{array}{l}
x_{n+1}=x_n \left(1+\Delta t  \left(r_1-d_1 x_n+y_n \left(\frac{\alpha _{12}}{h_2+y_n}-\frac{\beta _1}{e_1+x_n}\right)\right)\right)\\ \\

y_{n+1}=y_n \left(1+\Delta t  \left(r_2-d_2 y_n+x_n \left(\frac{\alpha _{21}}{h_1+x_n}-\frac{\beta _2}{e_2+y_n}\right)\right)\right) \\ \\

x_0>0, y_0>0. \\

\end{array}
\right.
\end{equation}
It is worth noting that the above system is not unconditionally positive due to the presence of negative terms that can allow the generation of negative solutions. This feature is considered to be precursor for numerical instabilities. 
For the calculation of the fixed points, the following system must have resolved:
\begin{equation}\label{eq:eul_fp}
\left\{
\begin{array}{l}
 x_{n}=x_n \left(1+\Delta t  \left(r_1-d_1 x_n+y_n \left(\frac{\alpha _{12}}{h_2+y_n}-\frac{\beta _1}{e_1+x_n}\right)\right)\right)\\ \\

 y_{n}=y_n \left(1+\Delta t  \left(r_2-d_2 y_n+x_n \left(\frac{\alpha _{21}}{h_1+x_n}-\frac{\beta _2}{e_2+y_n}\right)\right)\right)\\ \\

x_0>0,\, y_0>0. \\
\end{array}
\right.
\end{equation}
The scheme \eqref{eq:mutualism_timedisc_sol_eul} and the solutions coming from system \eqref{eq:eul_fp} will be compared in the Section \ref{sec:num_sim} with the results obtained by using NSFD scheme that will be shown in the following paragraph. 

\paragraph{NSFD scheme.} Now, we transform the differential equations in system \eqref{eq:mutualism_timecontinuos} respecting the following NSFD rules:
\begin{itemize}
\item II rule: denominators of left--hand side are replaced with $\phi=\phi(\Delta t)$;  
\item III rule and PESN criterion: the terms on the right of the equations are replaced with a non--local approximation. 
\end{itemize}
Then, the resulting finite difference system is:

\begin{equation}\label{eq:mutualism_timedisc}
\left\{
\begin{array}{l}
\frac{x_{n+1}-x_n}{\phi(\Delta t)}=r_{1}x_n+\alpha_{12}\frac{y_nx_n}{h_2+y_n}-\beta_1\frac{y_nx_{n+1}}{e_1+x_n}-d_1x_nx_{n+1}\\ \\

\frac{y_{n+1}-y_n}{\phi(\Delta t)}=r_2y_{n}+\alpha_{21}\frac{y_nx_n}{h_1+x_n}-\beta_2\frac{y_{n+1}x_n}{e_2+y_n}-d_2y_ny_{n+1}\\ \\

x_0>0, y_0>0. \\
\end{array}
\right.
\end{equation}
Then they are calculated $x_{n+1}$ and $y_{n+1}$:
\begin{equation}\label{eq:mutualism_timedisc1}
\left\{
\begin{array}{l}
 x_{n+1}=F_1(x_n,y_n,\phi,k_1)\\ \\

 y_{n+1}=F_2(x_n,y_n,\phi,k_2)\\ \\

x_0>0,\, y_0>0, \\
\end{array}
\right.
\end{equation}
where:

\begin{equation}\label{eq:F1}
F_1(x_n,y_n,\phi,k_1)=\frac{(e_1+x_n)(h_2+y_n)+\phi(e_1+x_n)(r_1(h_2+y_n)+y_n\alpha_{12})}{(e_1+x_n)(h_2+y_n)+\phi(h_2+y_n)(d_1x_n(e_1+x_n)+y_n\beta_1)},\\ \\
\end{equation}
\begin{equation}\label{eq:F2}
F_2(x_n,y_n,\phi,k_2)=\frac{(e_2+y_n)(h_1+x_n)+\phi(e_2+y_n)(r_2(h_1+x_n)+x_n\alpha_{21})}{(e_2+y_n)(h_1+x_n)+\phi(h_1+x_n)(d_2y_n(e_2+y_n)+x_n\beta_2)},\\
\end{equation}
\normalsize
in which $k_1$ and $k_2$ are the parameters already defined in Subsection \ref{subsec:ctd}. Since all components of $k_1$ and $k_2$ are non--negative we have $x_{n+1},y_{n+1}\ge 0$ for all $n\ge 0$, guaranteeing the respect of the positiveness condition.
\\Introducing $\Omega=\bigcup_{\tilde\xi\in\Gamma_d}\sigma(\tilde J)$, $\phi$ is given by the following expression \cite{mickens00}:
\begin{equation}\label{eq:phi}
\phi(\Delta t)=\frac{1-e^{-q\Delta t}}{q},
\end{equation}
in which:
\begin{equation}\label{eq:condition}
q\ge\max_{\Omega}\Big\{\frac{\lambda^2}{2|\Re(\lambda)|}\Big\}\,\,\,\rm if\,\,\,  \Re(\lambda)\neq 0\,\, for    \,\,\,\lambda\in\Omega.
\end{equation}
Following the definition given in \eqref{eq:fixed_point_disc}, the steady--states are calculated by solving the below system:
\begin{equation}\label{eq:mutualism_timedisc1_fp}
\left\{
\begin{array}{l}
F_1(x_n,y_n,\phi,k_1)=x_n\\ \\

F_2(x_n,y_n,\phi,k_2)=y_n.\\
\end{array}
\right.
\end{equation}
To discuss the stability of all ${\tilde\xi_n\in\Gamma_d}$, it is convenient to check if the Jury condition mentioned in eq. \eqref{eq:Jury} it is respected for each of these. To do this, it is need to calculate the Jacobian matrix $J_F(x_n,y_n)=(j_{ij})_{2\times 2}$. The elements of matrix are the following:

\begin{equation*} 
j_{11}(x_n,y_n)=\frac{(1+\phi r_1)(h_2+y_n)+\phi y_n\alpha_{12}\big((e_1+x_n)^2+\phi y_n\beta_1 (e_1+2x_n)\big)}{(h_2+y_n)((e_1+x_n)(1+\phi d_1 x_n)+\phi y_n\beta_1)^2},
\end{equation*}
\begin{align}j_{12}(x_n,y_n)=\frac{\phi x_n(e_1+x_n)\big(h_2(e_1+x_n)(1+\phi d_1 x_n)\alpha_{12}-\big((1+\phi r_1)(h_2+y_n)^2+\phi y_n^2\alpha_{12}\big)\beta_1\big)}{(h_2+y_n)^2((e_1+x_n)(1+\phi d_1 x_n)+\phi y_n\beta_1)^2},\end{align}

\begin{align*}j_{21}(x_n,y_n)=\frac{\phi y_n(e_2+y_n)\big(h_1(e_2+y_n)(1+\phi d_2 y_n)\alpha_{21}-\big((1+\phi r_2)(h_1+x_n)^2+\phi x_n^2\alpha_{21}\big)\beta_2\big)}{(h_1+x_n)^2((e_2+y_n)(1+\phi d_2 y_n)+\phi x_n\beta_2)^2},\end{align*}
\begin{align*}
j_{22}(x_n,y_n)=\frac{(1+\phi r_2)(h_1+x_n)+\phi x_n\alpha_{21}\big((e_2+y_n)^2+\phi x_n\beta_2 (e_2+2y_n)\big)}{(h_1+x_n)((e_2+y_n)(1+\phi d_2 y_n)+\phi x_n\beta_2)^2}.
\end{align*}
\normalsize

\section{Numerical Simulations}\label{sec:num_sim}

In this Section are presented some numerical simulations to test the dynamic consistence of the NSFD and the Forward--Euler schemes found in Subsection \ref{sub:dtd} with the continuous--time one described by the system \eqref{eq:mutualism_timecontinuos}.

\subsection{Steady--states analysis}

\begin{figure}%[htbp]
\centering
\includegraphics[width=0.95\textwidth]{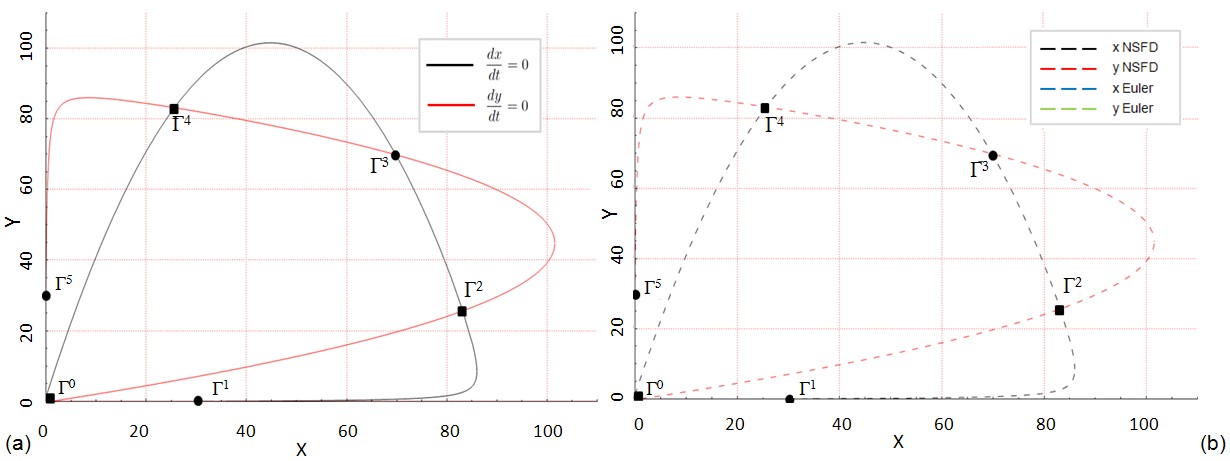}
\caption{\footnotesize{Comparison between phase--plane diagrams for continuous--time and discrete--time cases. (a) Zero--growth isoclines for system of eq. \eqref{eq:mutualism_timecontinuos} are represented as black ($dx/dt=0$) and red ($dy/dt=0$) lines. (b) Zero--growth isoclines come from the system \eqref{eq:mutualism_timedisc1_fp}: black dotted line for first equation and red dotted line for the second equation; zero--growth isoclines come from system \eqref{eq:eul_fp}: in blue the first equation and in green the second one (completely superimposed to the NSFD lines). For the discretization parameters, the following values have been used: $\Delta t=10^{-3}$, $q=10$. The steady--states are marked with letters $\Gamma^0$ to $\Gamma^5$: the squares ($\Gamma^0,\,\Gamma^2,\,\Gamma^4$) indicate the repellers of the system as well as the circles ($\Gamma^1,\,\Gamma^3,\,\Gamma^5$) indicate the attractors. The value assigned at each parameter is \cite{Holland_2005}: $r_1=r_2=0.3$, $\alpha_{12}=\alpha_{21}=0.6$, $q_1=q_2=1$, $\beta_1=\beta_2=0.2$, $d_1=d_2=0.01$, $h_1=h_2=0.3$ and $e_1=e_2=0.3$.}}\label{fig:phaseplane.jpg}
\end{figure}

In continuous--time domain the problem is largely discussed in \cite{Holland_deangelis2010} and \cite{Holland_2005}: these authors have performed many simulations assuming different values for the model's parameters, i.e., assuming different types of mutualism. In this paper a unique simulation for steady--state analysis is reported, hypothesizing for the involved parameters $r_1,\,r_2$, $\alpha_{12}$, $\alpha_{21}$, $\beta_{1}$, $\beta_2$, $e_1$, $e_2$, $d_1$ and $d_2$ the same values used by Holland and DeAngelis \cite{Holland_2005} for bi--directional mutualism. The steady--states have been calculated applying the condition \eqref{eq:fixed_point_cont} to the system of differential equation \eqref{eq:mutualism_timecontinuos}. From graphical point of view, the steady--states are found plotting the zero--growth isoclines of each system's equation and are represented by the points where the isoclines intersect between them. The phase--plane diagrams for the $x$--$y$ dynamics is showed in \figurename~\ref{fig:phaseplane.jpg}\textit{a} where the zero--growth isoclines are depicted. 

\begin{table}%[htbp]
\footnotesize
\begin{center}\label{tab:1}
  \begin{tabular}{@{\extracolsep{\fill}} | c | c | c | c | c | c | }
  \hline
  \textbf{Steady--state} & \textbf{x--coordinate} & \textbf{y--coordinate} & \textbf{$\lambda_1$} & \textbf{$\lambda_2$} & \textbf{Stability} \\
  \hline
  $\Gamma^0$  & 0  & 0  & 0.30 & 0.30 & R \\
  \hline
  $\Gamma^1$  & 30  & 0  & -19.11 & -0.30 & A \\
  \hline
  $\Gamma^2$  & 25.62  & 83.17  & -0.80 &  0.41& R  \\
  \hline
  $\Gamma^3$  & 69.83  & 69.83  & -0.70 & -0.30 & A  \\
  \hline
  $\Gamma^4$  & 83.17  & 25.62  & -0.80 & 0.41 & R  \\
  \hline
  $\Gamma^5$  & 0  & 30  & -19.11 & -0.30 & A  \\
  \hline
  \end{tabular}
\caption{Steady--states of \figurename~\ref{fig:phaseplane.jpg}. A=attractor, R=repeller.}
\end{center}
\end{table}

\begin{figure}%[htbp]
\centering
\includegraphics[width=0.95\textwidth]{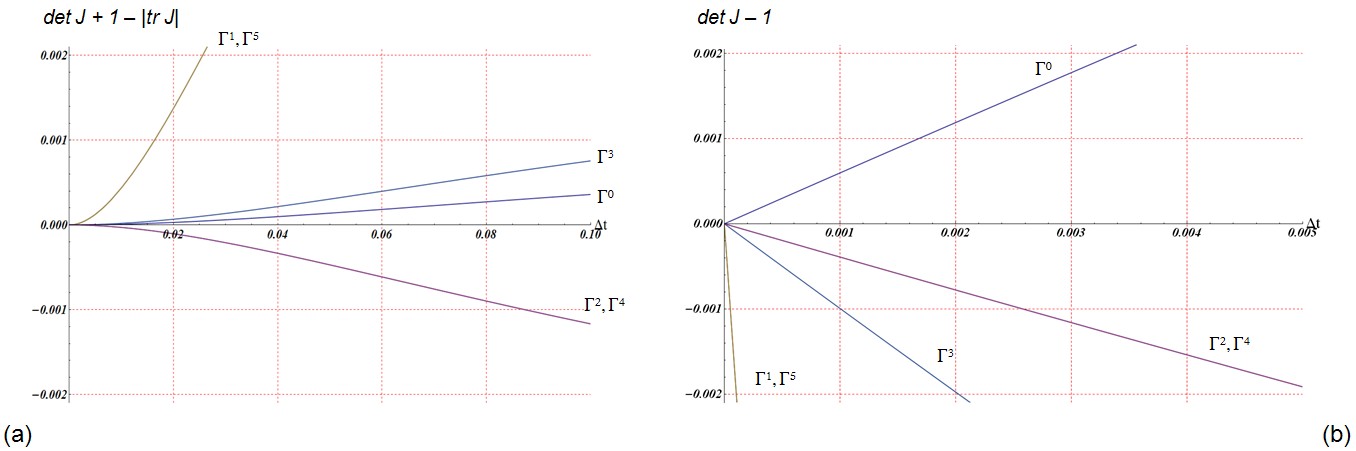}
\caption{\footnotesize{Jury condition validation for NSFD scheme. (a) Plot of eq. \eqref{eq:det-1} versus the time step $\Delta t$. Each curve is related to at least one of the steady--states $\Gamma^0,\,\Gamma^1,...,\Gamma^5$. (b) Plot of eq. \eqref{eq:det+1-tr} versus $\Delta t$.}}\label{fig:jury}
\end{figure}

In the discrete case, NSFD derived scheme has been compared with the Euler one calculating the solutions of the systems \eqref{eq:mutualism_timedisc1_fp} and \eqref{eq:eul_fp} respectively. The \figurename~\ref{fig:phaseplane.jpg}\textit{b} represents the isoclines described by the Euler and NSFD schemes for $\Delta t=10^{-3}$. From those graphs can be seen that, for the assumed parameters, the dynamical system has 6 steady--states, indicated with $\Gamma^0$, $\Gamma^1$,.., $\Gamma^5$ in counterclockwise sense. Also, none of the two discretization methods introduces spurious solutions: the coordinates of the steady--states are the same and the isoclined are perfectly superimposed. The Table 1 summarizes the results of these simulations and the conclusions on steady--states stability for continuous and discrete--time schemes. From the examination of the signs of the eigenvalues, it emerges that $\Gamma^0$, $\Gamma^2$ and $\Gamma^4$ are repellers (indicated in figure with a black square) and $\Gamma^1$, $\Gamma^3$ and $\Gamma^5$ are attractors (black circle). 

However, when $\Delta t$ is increased up to $8.445\cdot 10^{-2}$ we observe that the Euler scheme converges toward wrong steady--states or the two populations have a boundless growth. On the contrary the phase--plane of the NSFD remains unchanged. 

Since one of the requisites for the dynamic consistence is the invariance of the stability conditions regardless to the discretization step--size, has been checked for NSFD scheme if the Jury condition is guaranteed or not independently from the value of $\Delta t$. This check is not performed for Euler scheme because, as explained above, it fails the convergence to the correct fixed points for certain values of the step--size. 

In order to verify this, the condition \eqref{eq:Jury} is splitted in two inequalities that will be proved for all $\tilde\xi\in\Gamma_{d}$, making explicit the dependence of the Jacobian elements from $\Delta t$: 

\begin{equation}\label{eq:det-1}
\det (\tilde J(\Delta t))-1<0,
\end{equation}
\begin{equation}\label{eq:det+1-tr}
\det (\tilde J(\Delta t))+1-\big|\operatorname{tr}(\tilde J(\Delta t))\big|>0.
\end{equation}
The Figures \ref{fig:jury}\textit{a} and \ref{fig:jury}\textit{b} represent respectively the left term of eq. \eqref{eq:det-1} and \eqref{eq:det+1-tr}, both evaluated as function of $\Delta t$: in both graphs every curve is related to at least one steady--state $\Gamma^0,\,\Gamma^1,...,\Gamma^5$. Only for reasons of correct graphic visualization the two figures have different scales on abscissa. It is easy to see that both the inequalities \eqref{eq:det-1} and \eqref{eq:det+1-tr} are satisfied for NSFD schemes only from $\Gamma^1$, $\Gamma^3$ and $\Gamma^5$ steady--states for each value of $\Delta t$ represented. Nevertheless, has been checked that all the curves are monotone for $0<\Delta t\le 10$, then the conclusions about the steady--states stability are unchanged up to $\Delta t=10$.

\subsection{Transient dynamics}

In this Section a transient analysis has been performed with the purpose to emphasize the possible criticalities of the two considered discretization schemes. 

\begin{figure}[htbp]
\centering
\includegraphics[width=0.8\textwidth]{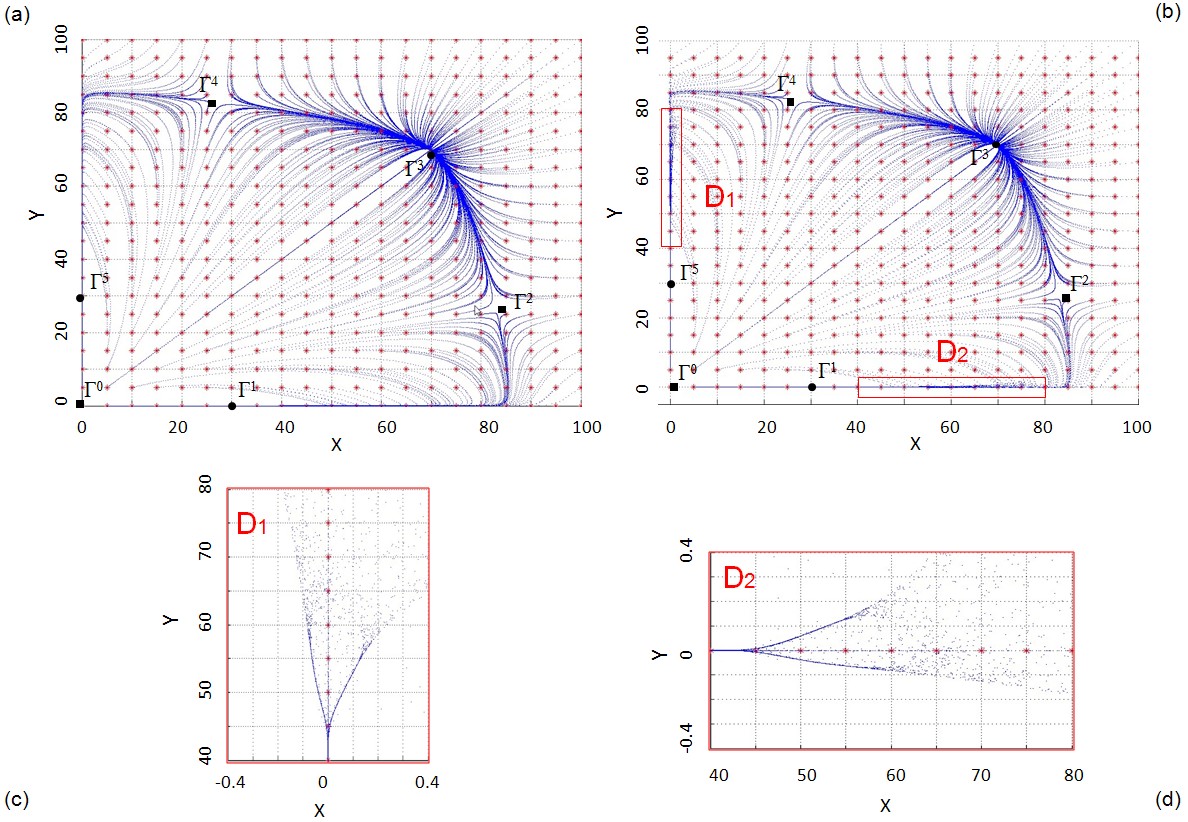}
\caption{\footnotesize{(a) Phase--plane portrait of system \eqref{eq:mutualism_timedisc1}. (b) Phase--plane portrait of system \eqref{eq:mutualism_timedisc_sol_eul}. (c) Particular of subdomain $\rm D_1$ belongs to (b). (d) Particular of subdomain $\rm D_2$ belongs to (b). The same parameters adopted for \figurename~\ref{fig:phaseplane.jpg} computation have been used, with the exception of $\Delta t=6\cdot 10^{-2}$.}}\label{fig:phase-portrait.jpg}
\end{figure}

The phase--plane portraits of NSFD and Forward--Euler difference equations has been shown in \figurename~\ref{fig:phase-portrait.jpg}\textit{a} and \textit{b} respectively: all the used parameters are the same used for computation of \figurename~\ref{fig:phaseplane.jpg}, exception for the time step that now is $\Delta t=6\cdot 10^{-2}$. From the comparison it is possible to observe some differences highlighted by red rectangles in the \figurename~\ref{fig:phase-portrait.jpg}\textit{b} named with $\rm D_1$ and $\rm D_2$. The subplots \ref{fig:phase-portrait.jpg}\textit{c} and \textit{d} represent the magnification of what happens in $\rm D_1$ and $\rm D_2$ respectively: it is possible to note that in subdomain $\rm D_1$ the population $x$ assumes negative values for about $y\ge 45$ and the population y assumes negative values for about $x\ge 45$. All these negative values appear during the transient.

\begin{figure}[htbp]
\centering
\includegraphics[width=0.8\textwidth]{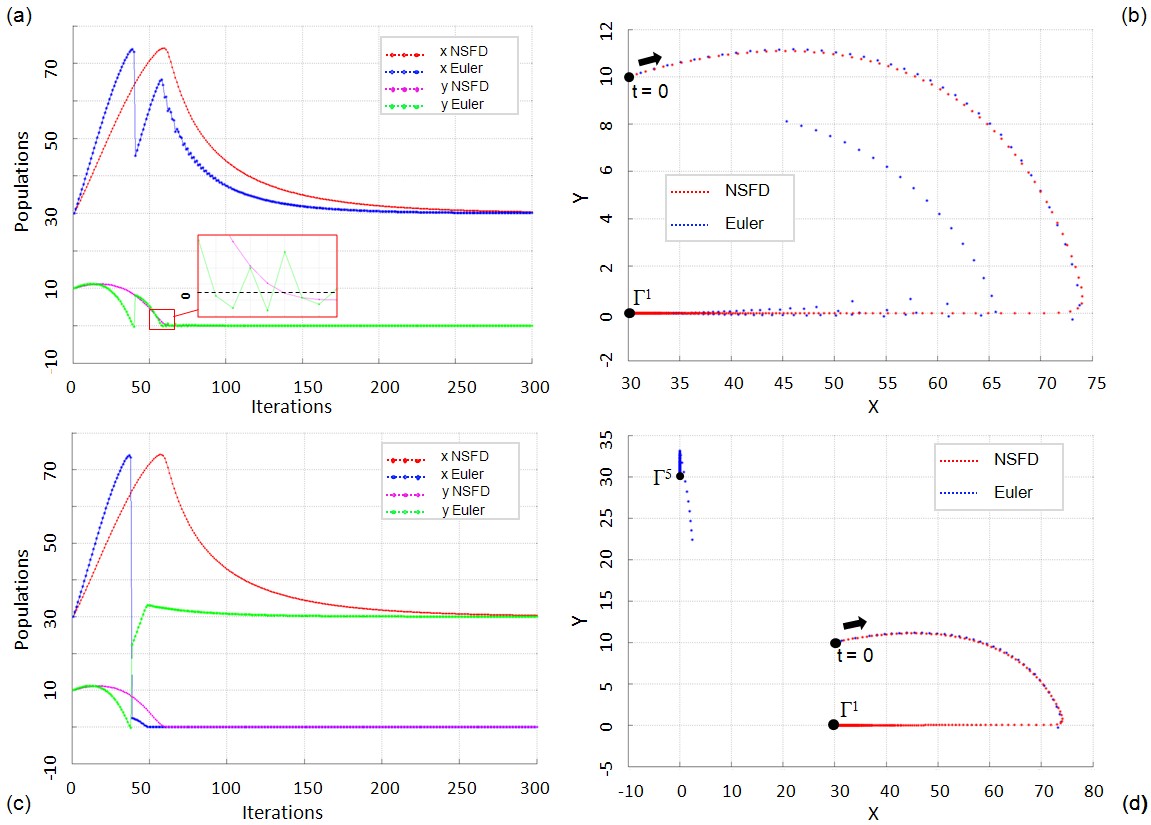}
\caption{\footnotesize{Comparison between transients of Euler, eqs. \eqref{eq:mutualism_timedisc_sol_eul}, and NSFD, eqs. \eqref{eq:mutualism_timedisc1}, difference equations. (a) $x$ populations \textit{vs} iteration number: red points for NSFD and blue points for Euler scheme; $y$ populations \textit{vs} iteration number: violet points
for NSFD and green points for Euler scheme. $\Delta t=10^{-3}$. (b) Phase--plane of (a). (c) The same calculus of plot (a), but assuming $\Delta t=8.44\cdot 10^{-2}$. (d) Phase--plane of (c).}}\label{fig:eulnsfd_transit.jpg}
\end{figure}

The Figures \ref{fig:eulnsfd_transit.jpg}\textit{a} and \textit{c} show a plot of the populations \textit{vs} the iteration number using respectively \eqref{eq:mutualism_timedisc_sol_eul} and \eqref{eq:mutualism_timedisc1} schemes. In order to show as the numerical instabilities arise for larger values of $\Delta t$ a comparison between Euler and NSFD schemes has been performed by assuming respectively $\Delta t=10^{-3}$ (see Figure \ref{fig:eulnsfd_transit.jpg}\textit{a}) and $8.44\cdot 10^{-2}$ (see Figure \ref{fig:eulnsfd_transit.jpg}\textit{c}) with initial populations $(x_0,y_0)=(30,10)$. It can be seen from the Figure \ref{fig:eulnsfd_transit.jpg}\textit{a} that for both Euler and NSFD equations, after a transient in which the populations $x$ and $y$ reach their maximum values, about $72$ and $11$ individuals, they converge smoothly toward $x=30$ and $y=0$ individuals (steady--state $\Gamma^1$). The relative phase--plane is depicted in the \figurename~\ref{fig:eulnsfd_transit.jpg}\textit{b}, where the two trajectories are perfectly superimposed. By increasing the time step up to $\Delta t=8.44\cdot 10^{-2}$ the populations behave as in \figurename~\ref{fig:eulnsfd_transit.jpg}\textit{c}, calculated keeping all other parameters unchanged. The remarkable fact is that, while for NSFD the transient preserves the same aspect of \figurename~\ref{fig:eulnsfd_transit.jpg}\textit{a}, the Euler scheme shows jumps for both $x$ and $y$ populations around the 40th iteration and an anomalous behavior that also continues in later iterations. In \figurename~\ref{fig:eulnsfd_transit.jpg}\textit{d} is plotted the relative phase--plane where the difference between NSFD and Euler trajectories are particularly evident; also note the difference of the Euler trajectories between the subplots \textit{b} and \textit{d}.

From the tests that have been performed it is possible to infer that the first numerical instabilities in Forward--Euler scheme occur when $\Delta t$ overcomes the critical value of $\Delta t_c\simeq 5\cdot 10^{-2}$. The reason is that in Forward--Euler the step sizes larger than $\Delta t_c$ are greater than some characteristic time relevant of this physical problem. The same is not true employing the NSFD technique, in which the numerical instabilities are eliminated since the selected step size of eq. \eqref{eq:phi} is never larger than the smallest time scale of the system.

\section{Conclusions}
On the base of the work published by Holland and DeAngelis in 2009, in which is constructed and analyzed a population dynamic continuous--time model linking the consumer functional responses of one mutualistic species with resources supplied by another, in this paper a strong nonstandard finite difference scheme to solve the continuous--time model is constructed. As shown by numerical simulations, the NSFD framework reproduces the dynamical features of the continuous--time analogous. In particular, it is proved that the method guarantees the correct asymptotic behavior regardless from the size of the time step. It was shown that, with the same initial conditions, the Forward--Euler induces scheme--dependent numerical instabilities for $\Delta t$ greater than a critical values, that manifest as jumps in the population evolution in time domain and convergence toward wrong steady--state solutions. On the other hand, in tests that were performed, the NSFD scheme shows none of these problems, confirming the versatility of the method.

\bibliographystyle{plain}

\label{lastpage}
\end{document}